# Investigating the Effects of Mobility Metrics in Mobile Ad Hoc Networks

Mohsin Ur Rahman*, Aftab Alam†, and Zia ur Rahman§
*†§Department of Computer Science, University of Malakand
Email: *mohsin@uom.edu.pk, †aalab@uom.edu.pk, §ziacs@uom.edu.pk

*Abstract*—Mobile Ad Hoc Networks (MANETs) are formed by a collection of mobile nodes (MNs) that are capable of moving from one location to another location. These networks are widely identified by their unique characteristics such as lack of infrastructure, mobility and multi-hop communication. Unlike traditional (wired) networks, MNs in MANETs do not rely on any infrastructure or central management. Mobility allows MNs to move at different values of speed. Multi-hop communication is used to deliver data across the entire network. Due to mobility and changing network topology, the performance of these networks is significantly affected by the choice of mobility models and routing protocols. Our research work aims to evaluate the effects of mobility metrics on distinguishing between entity and group mobility models in MANETs. In addition, we demonstrate the interactions between mobility metrics and performance metrics. We also investigate how effective are the mobility metrics and which metrics can clearly distinguish between entity and group mobility models?

We performed extensive simulations using network simulator, ns-2.35, to capture mobility metrics as well as different performance metrics. Simulation results reveal the efficiency of mobility metrics on distinguishing between entity and group mobility models. We also obtain useful interactions between mobility metrics and performance metrics. The results presented in this paper provide new insights into the variability of mobility metrics in MANETs. Furthermore, our simulation results reveal better understanding of the relationships between mobility metrics (e.g., relative speed, node degree, network partitions etc.) and performance metrics (e.g., packet delivery ratio, end-to-end delay and normalized routing load).

## I. INTRODUCTION

Mobile Ad Hoc Networks (MANETs) represent distributed networks formed by a collection of mobile nodes (MNs). These networks are widely identified by their unique characteristics such as lack of infrastructure, mobility and multi-hop communication. Unlike traditional (wired) networks, MNs in MANETs do not rely on any infrastructure or central management. Due to these characteristics MANETs are susceptible to a wide range of attacks such as Denial of Service (DoS) [1], Sybil and replication attacks [2]. MANETs are commonly used in various real-life environments such as search and rescue operations, robot networks, battlefield communication, natural catastrophes and tracking and surveillance operations [3].

To advance the field of MANETs, researchers have proposed a variety of routing protocols, covering all the layers in the protocol stack. There are two ways to evaluate the performance of these protocols; *real test-bed experiments* and *simulations*. The former method is expensive and time consuming because it requires costly devices and extraction of real-world traces. The latter is by far the widely used method for designing and evaluating the performance of MANET routing protocols [?].

A Mobility model serve as a backbone when performing simulations in MANETs. It is used to mimic the movement behavior/patterns of real-world objects such as the movement of vehicles, robots or people in real-life environments. In addition, the choice of a mobility model can greatly alter the performance of routing protocols. For example, Bai *et al.* [4] performed extensive simulations and found that the performance of protocols largely depends on the underlying mobility models.

Various mobility models have been designed to improve data delivery in different movement scenarios. These models are used to imitate the moving patterns of real-world devices. For instance, people with mobile devices move to different places based on their social connections. Similarly, other models are used to mimic users' movement in battlefields, cities, urban areas, grids and so on. These models are broadly categorized into: (i) entity mobility models and (ii) group mobility models. The former are used to mimic the arbitrary movements of individual nodes. Random Waypoint (RWP) model is an example of entity model. In contrast, group mobility models are designed to depict the movement patterns of a group(s) of MNs such as people moving in groups, a group of robots performing different task and many more. Reference Point Group (RPGM) is a famous group mobility model proposed by Hong et al. [5], [6].

Current studies covering the performance of MANET mobility models and routing protocols have some limitations. For example, many of the previous studies evaluated the effects of speed [7]–[11]. A few assessed the influence of node density. Others have performed simulation for different values of transmission range [11]–[13]. Thus, there is plenty of gap for assessing the impacts of other conditions such as the interactions between mobility metrics and performance metrics and the efficiency of mobility metrics on classifying the same or different types of models.

The rest of the paper is organized as follows. Related work is discussed in section II. MANETs mobility metrics are discussed in section III. In Section IV, we discuss MANETs mobility models. In section V, we describe our simulation configuration and input parameters for the different mobility models. Simulation results are discussed in section VI. Finally, section VII concludes the paper.

## II. RELATED WORK

Bai *et al.* [4] performed extensive simulations and proposed a framework for evaluating the impacts of mobility patterns. They evaluated the effects of mobility on the performance of different protocols. Moreover, they also validated the effects of mobility metrics on the performance of AODV, DSR and DSDV protocols.

The authors in [14] carried out our simulations for five different terrain sizes with respect to various pause times and observed that DSR outperforms AODV for small terrain areas in terms of packet loss, delay and PDF. An efficient algorithm usually has low packet loss and, for these areas, DSR satisfies this property. The packet delivery ratio is less than 100% due to hidden and exposed terminal phenomenon in MANET. Medium terrain areas have low node density and increased chances of link-breakages compared to small areas. Hence, AODV provides the best performance for medium terrain areas. For such areas, AODV performs quite predictably, delivering more than 99% packets with negligible delay and packet loss. Finally, when simulated in high terrain areas, AODV outperformed DSR in terms of delay. Furthermore, both protocols drop nearly the same amount of packets when simulated in such larger areas.

Theoleyre et al. [15] investigated the influence of node density on link duration (LD) over different mobility models. They observed that LD does not impact most mobility models except for group-based models, i.e., RPGM and NCMM. Thus, Random, i.e., entity mobility models were not affected at different values of node density.

The authors in [16] used the method of linear regression in order to design models that are capable of predicting the number of network partitions and path size from the input parameters. They proposed different models, which were suitable for prediction in square and rectangular scenarios. The authors claims that speed and pause time produce little effect on the number of network partition (NP) metric. However, the investigations in this study are limited only to the Random Waypoint model.

The author in [17] evaluated node degree (ND) and showed its effects on node density, size of the simulation area and transmission range. However, they only used Random Waypoint model. Simulation results show that ND linearly increases as the number of nodes are increased. They also considered square scenarios for simulations and observed that the values of X and Y causes an exponential decay in ND.

The author in [18] investigated the relationships between mobility metrics and the performance of routing protocols. Simulation results reveal that Link Duration (LD) and Path duration (PD) are linearly associated with the performance metrics in terms of routing overhead and throughput. However, this study is only limited to two performance metrics.

It is evident from the literature review that mobility metrics have a direct impact on the performance metrics, thus it seems suitable to investigate the relationships between them. Therefore, the basic objectives of this research are to investigate the following research questions.

- *RQ1:* How effective are these metrics on distinguishing mobility models?
- *RQ2:* Which mobility metrics are strongly related to which performance metrics?
- *RQ3:* Which metrics can clearly distinguish between entity and group mobility model?

## III. MOBILITY METRICS

Mobility metrics are protocol independent features used to assess the behavior of mobility models. We evaluated these metrics based on the following assumptions:

1) The same number of nodes were used for obtaining each metric.
2) The transmission range was pre-defined and same for each run of the simulation
3) Communication was bidirectional between every pair of MNs
4) we considered scenarios with two-dimensional geometry

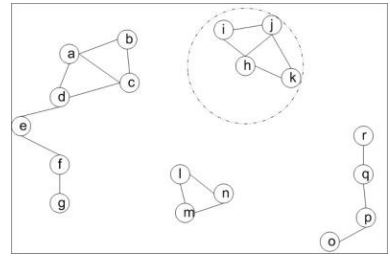

Fig. 1. Illustration of MANET

### A. Node Degree (ND)

We can simply define node degree as the total number of nodes located within the radio range of a node. For example, ND for node h is 3 (see Fig. 1) since only 3 nodes are located in its radio range. This metric is a subject of extensive research in MANETs. For instance, the authors in [17] used Random Waypoint (RWP) model and evaluated the impacts of ND and radio range.

### B. Number of Network Partitions (NP)

This metric is used to provide information about the number of partitions in a network. As shown in Fig. 1, the value of NP is 4 because the given network consists of four partitions. This metric is also a hot subject for researchers. For instance, the authors in [19] explored the number of partitions by conducting simulations for large-scale networks in MANETs.

### C. Link Changes (LC)

LC is used to reveal information about the communication status between two nodes. For example, link may be up or down showing the status of communication. This metric was initially proposed to differentiate the movements patterns of Reference Point Group (RPGM) and Random Waypoint (RWP) models [20].

## D. Link Duration (LD)

As the name suggests, LD is used to reveal information about the total "duration" of time when two MNs communicate with each other. This metric is widely used in MANET research. The authors in [21] used this metric and showed that LD can be used to gauge the performance of routing protocols.

## E. Relative Speed (RS)

This metrics is related to the speed of MNs. For instance, it can be used to indicate the difference between the speed of two MNs, i.e., $j$ and $k$ at time slot $t$.

$$[v_j(t) - v_k(t)]$$

## IV. MANETs MOBILITY MODELS

A mobility model is a process that requires input parameters in order to produce output mobility scenarios/traces as shown in Fig. 2. The output traces can be used to obtain useful statistical information about mobility metrics.

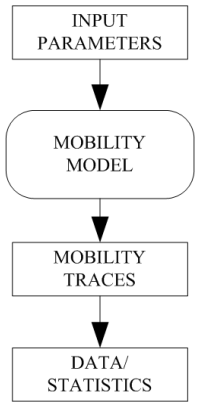

Fig. 2. Mobility models as an input/output process

Mobility models are developed by assuming mathematical computations in order to predict the walking patterns of real-world objects. For example, computation of the next values of speed and direction. However, MNs in MANETs have limited capabilities such as limited battery life, memory and processing power. Therefore, both mobility models and routing protocols should be carefully designed to achieve optimum performance. These models are broadly categorized into two classes: *(i) Entity mobility models* and *(ii) Group mobility models*. In the former, each MN continues its movement in a completely independent fashion, i.e., independent of its previous movement. In the latter, MNs are placed in groups and continue their motion following their "group leader" [22] [23]. The mobility models considered in this research work are briefly discussed in the following section.

## A. Radom Waypoint (RWP) Model

RWP [24] is the simplest model because it only requires a pre-defined speed and pause time to simulate a network scenario. This model is widely used when evaluating the performance of MANET routing protocols. The basic concept of this model is that MNs initially select random points in the simulation area and then start moving towards it with a pre-defined speed. Upon reaching that points, they remain silent for a pre-defined amount of time. After that, they select new random points and repeat the said process again. Thus, RWP only requires two input parameters for simulations, i.e., *speed*, and *pause* time. This model is extensively investigated in MANETs environments [25].

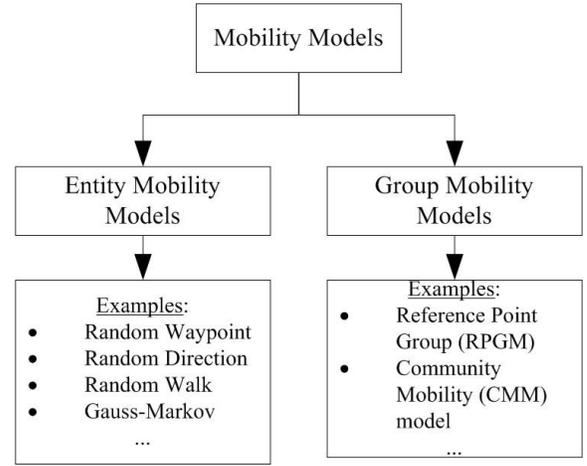

Fig. 3. Classification of mobility models in MANETs

## B. Gauss Markov Mobility Model (GM)

In this model, one tuning parameters $a$ is used to adapt to different level of randomness. Each MN is allocated a current speed and direction. Furthermore, movements occur in fixed time intervals. The values of speed and direction are updated after each time interval, $t$ [25]. At $j^{th}$ time interval, the current values of speed and direction of a node that depend on it's previous $(j-1)^{th}$ time interval are calculated by using the following equations:

$$v_j = av_{j-1} + (1-a)\bar{v} + \sqrt{1-(a^2)}v_{x_{j-1}}$$
$$d_j = ad_{j-1} + (1-a)\bar{d} + \sqrt{1-(a^2)}d_{x_{j-1}} \quad (1)$$

Where $v_j$ and $d_j$ are the new values of speed and direction at time intervals $j$, $a$ is a parameter (tuning) used to vary randomness of the GM movements, $\bar{v}$ and $\bar{d}$ are constant representing the mean values of velocity and direction.

At each time interval, the next location of each node is calculated by considering the previous values. For example, at at $j^{th}$ time interval, the next position can be calculated by using the following equations.

$$x_j = x_{j-1} + s_{j-1} \cos a_{j-1}$$
$$y_j = y_{j-1} + s_{j-1} \sin a_{j-1} \quad (2)$$

where $(x_j, x_{j-1})$ and $(y_j, y_{j-1})$ indicate the values of x and y coordinates at $j^{th}$ and $(j-1)^{th}$ time intervals, respectively. $s_{j-1}$ and $a_{j-1}$ represent the values of speed and direction at $(j-1)^{th}$ time interval.

## C. Nomadic Community Mobility Model (NCMM)

This model is commonly used to represent those scenarios where a group of MNs randomly moves to a specific location and then select another location and so on. The basic concept of this model is to represent the mobility patterns of people. For instance, people with *mobile devices* may move in groups or visit different locations depending upon their social relationships [26].

## D. Reference Point Group (RPGM) Mobility Model

This model is also a famous group-based model proposed in [20]. The basic concept of this model is based on the movement of a group of mobile nodes chasing their group leader. Specifically, MNs select a "logical center" or a group leader, which is responsible for adjusting the speed and direction of its members. Moreover, the group leader chooses a random point or "checkpoint" and starts moving towards it with a pre-defined speed. All the members of a group follow the path travelled by their group leader. Each group uses a motion path to calculate new destination point. This motion path is computed using "checkpoints" [27] [28].

## V. SIMULATION METHODOLOGY

We used BonnMotion [29] to generate different scenarios and statistics for RWP, GM, RPGM and Nomadic Community (NCMM) models. In all of these scenarios, 90 nodes are simulated in a simulation area of 1000m x 1000m. The duration of simulation is 900 sec. Group size for both group mobility models, i.e., RPGM and NCMM is set to 5. Furthermore, the maximum speed $s$ is set to 20 m/s. All the simulation results presented in this paper are averaged over 25 randomly generated scenarios.

In order to investigate the relationship between mobility metrics and performance metrics, we again performed simulations using Network Simulator, NS-2.35 [30]. The same set of parameters were used to generate scenarios, which were later exported to ns-2 to obtain values of PDR, delay and NRL as shown in Fig. 4(a,b,c). We simulated 90 nodes in a simulation area of $1000m^2$ for a duration of 300 seconds. The value of group size is five for both group mobility models. Thus both RPGM and NCMSS have 18 groups each containing five mobile nodes. We used CBR traffic sources to generate UDP traffic. Simulation parameters are shown in Table 1.

In order to ensure a fair comparison, it is essential to generate unbiased scenarios for each mobility model. Thus, we ensured a fair comparison by providing equal configurations for speed, transmission range, number of nodes and size of the simulation area. Values of input parameters for each mobility model are listed in Table 2.

| Parameters | Value(s) |
|---|---|
| Area Size | 1000m x 1000m |
| Simulation Time | 300 sec |
| Antenna | Omni |
| Transmission range | 250m |
| Mobility Models | RWP, GM, RPGM, NCMM |
| Group size (RPGM,NCMM) | 5 |
| Maximum Speed | 20 m/s |
| Maximum Pause Time | 10 sec |
| Application | CBR |
| Maximum Connections | 15 |
| Number of Nodes | 90 |
| Packet Size | 512 bytes |
| Sending rate | 4 packets/sec |
| Data Rate | 2 MBps |
| Protocol | AODV |
| MAC Protocol | MAC/802.11 |
| Scenario Generator | BonnMotion v2.1 |

TABLE I
SIMULATION PARAMETERS

| Parameters | RWP | GM | NCMM | RPGM |
|---|---|---|---|---|
| Number of Nodes | 90 | 90 | 90 | 90 |
| Maximum Speed | 20 | 20 | 20 | 20 |
| Pause Time | 10 | N/A | 10 | 10 |
| Radio Range | 75 | 75 | 75 | 75 |
| Area Lenght(X) | 1000 | 1000 | 1000 | 1000 |
| Area Width(Y) | 1000 | 1000 | 1000 | 1000 |
| Group Size | N/A | N/A | 5 | 5 |

TABLE II
VALUES OF INPUT PARAMETERS FOR EACH MOBILITY MODEL

## VI. RESULTS AND ANALYSIS

The effectiveness of average link duration (LD) to distinguish between entity and group models is shown in Fig. 5(a). It is obvious from the given results that group mobility models have higher LD values whereas random models the lowest. This is because MNs in groups move at speeds that differ by a small fraction from the group leader. We found a relationship between Normalized Routing Load (NRL) and LD. Models with higher values of LD show lowest NRL in the network as shown in Fig. 4(c). In other words, group mobility models minimize routing overhead by allowing nodes to communicate for a longer time as compared to random models.

The usefulness of the number of link changes (LC) on distinguishing between both classes of mobility models is shown in Fig. 5(b). The results show conflicting values for both types of models. For instance, LC values for RWP and RPGM are nearly the same, but NCMM is showing the highest values outperforming the other mobility models. Thus, our results also validate the low efficiency of LC to distinguish between both classes of models [31].

Fig. 5(c) shows the classification accuracy of node degree (ND). Simulation results indicate that ND can sufficiently distinguish between both types of modes. As it can be observed, group-based models show higher values of ND, while random models have the lowest. This happens because MNs in both RPGM and NCMM collectively move from one position to another, which implies that the probability of the number of nodes located within the transmission range is high. On the other hand, random models are designed by assuming random placement and movement in the simulation area, thus generating lowest values of ND. Furthermore, a relationship exists between ND and end-to-end delay as shown in Fig. 4(b). Thus, our results confirm that ND can clearly distinguish between both classes of models and produces a strong impact in terms of end-to-end delay in the network. However, only GM is showing conflicting values because it is designed by assuming minor changes in both speed and direction. Therefore, GM produces higher delay at faster speed conditions.

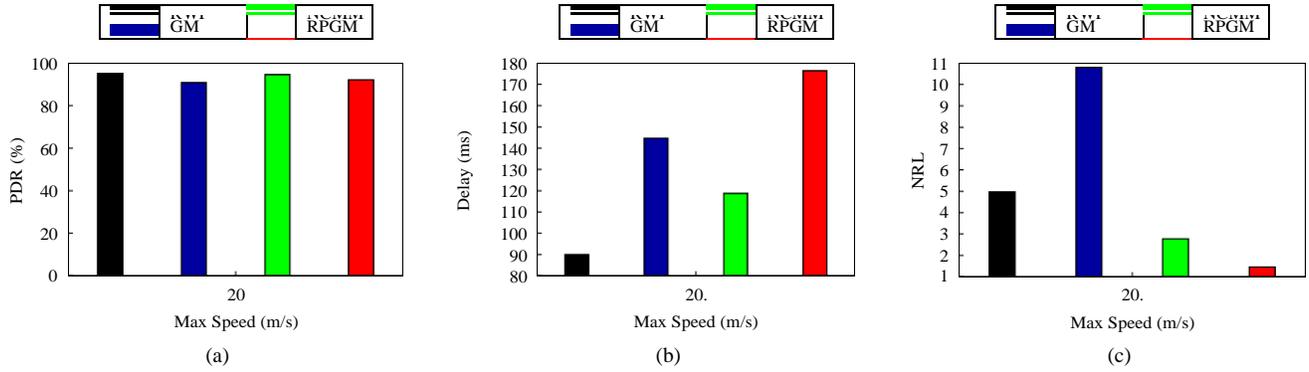

Fig. 4. Simulation results for performance metrics (a) PDR vs. max speed (b) Delay vs. max speed and (c) NRL vs. max speed (90 nodes)

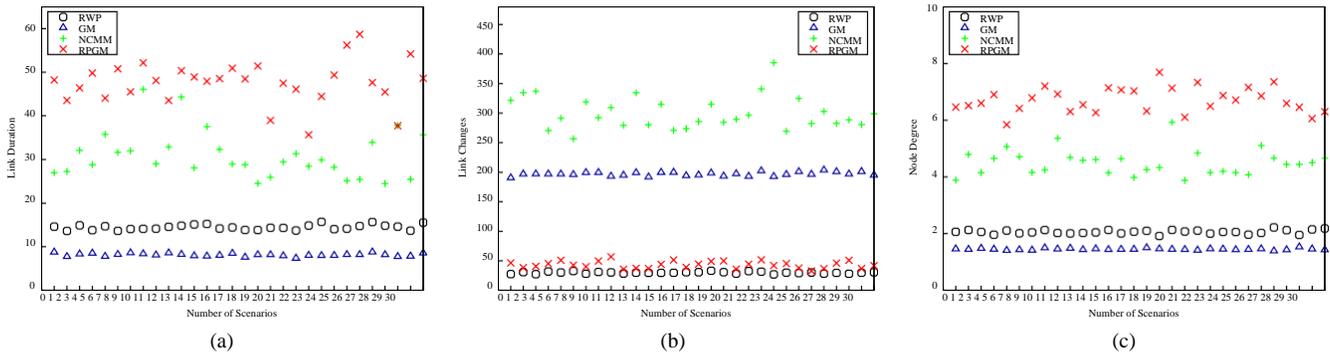

Fig. 5. Effectiveness of (a) relative speed (RS) (b) link changes (LC) and (c) node degree (ND) on distinguishing between entity and group mobility models

As shown in Fig. 6, number of network partitions (NP) also proves to be a good metric as the given results clearly separate random from group-based models. It is obvious from the given results that RPGM creates less partitions in the network. This metric affects routing load as shown in Fig. 4(c). For instance, RPGM and NCMM are showing less NRL as compared to both random models. It is important to observe that NCMM shows better PDR (%) by minimizing both NRL and end-to-end delay. Thus, less or moderate number of partitions seem to be a good design choice.

Finally, the usefulness of relative speed (RS) is shown in Fig. 7. This metric is also capable of discriminating different models. For example, RPGM shows lowest values of RS as compared to the other models. This is possibly because nodes in this model move at low proximity following their group leaders. However, RS is unable to correctly distinguish GM model because this model assumes minor changes in speed. Thus, GM is showing an opposite behavior because of the higher speed configuration. On the other hand, RS values for RWP are higher than the two group-based mobility models. However, this metric is not clearly distinguishing the same type of models. For example, RS values for RPGM and NCMM show great variation. Both entity models are also showing the same behavior.

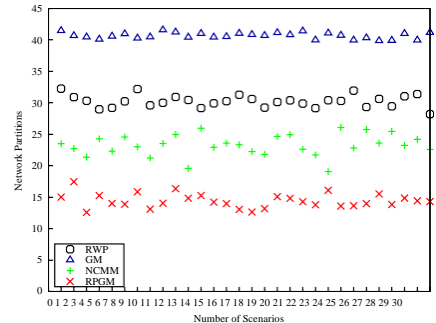

Fig. 6. Effectiveness of Network Partition (NP) on distinguishing between entity and group mobility models

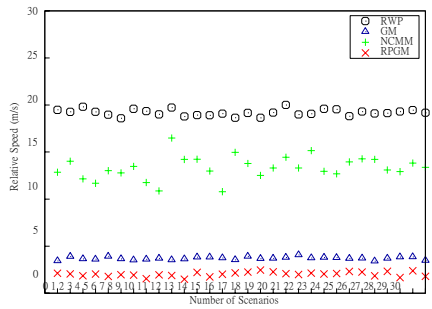

Fig. 7. Effectiveness of Relative Speed (RS) on distinguishing between entity and group mobility models

## VII. Conclusions

In this paper, we performed extensive simulations to analyze the effects of mobility metrics on distinguishing between entity and group mobility models with the following conclusions and answers to our three main research questions.

<u>Conclusion 1:</u> Our simulation results reveal that node degree (ND), link duration (LD) and network partitions (NP) metrics are highly effective for discriminating between both types of models. We also showed that the accuracy of these metrics depends on input parameters. For instance, we obtained different values for LC as compared to the findings previously reported. Our simulation results also corroborate the low efficiency of the LC metric (*RQ1*).

<u>Conclusion 2:</u> We observed that ND and LD can sufficiently distinguish between both types of models. Furthermore, both of these metric can greatly alter the performance of routing protocols in terms of end-to-end delay and normalized routing load, respectively. We obtained similar results for network partition metric (*RQ2*).

<u>Conclusion 3:</u> We observed that some metrics clearly distinguished between entity and group-based models. For example, random models produced lowest values of LD and ND as compared to group mobility models (*RQ3*).

<u>Conclusion 4:</u> We also investigated the influence of relative speed (RS) and showed that RS values for random models are the highest except for GM model. We also observed that NCMM shows better PDR (%) by minimizing both NRL and end-to end delay. Thus, this research reveals the importance of mobility metrics in distinguishing between both classes of model. Hence, mobility models should be designed by considering the effects of these metrics.

## Acknowledgment

The author would like to thank the members of the networking research group, University of Malakand, for providing valuable suggestions and comments.